\documentclass[amssymb,amsmath,prl,amsfonts,superscriptaddress,twocolumn,showpacs]{revtex4}

\usepackage{graphicx}
\usepackage{color} 
\usepackage{hyperref} 
\renewcommand{\vec}[1]{\mathbf{#1}}
\renewcommand{\d}{\mathrm{d}}

\definecolor{blue}{rgb}{0,0,0.76}

\usepackage{textcase}
\usepackage{titlesec}
\makeatletter
\newcommand*{\balancecolsandclearpage}{%
  \close@column@grid
  \clearpage
  \twocolumngrid
}
\makeatother
\usepackage[final]{pdfpages}

\begin{document}

\title{Braiding a flock: winding statistics of interacting flying spins}
\author{Jean-Baptiste Caussin}
\affiliation{Laboratoire de Physique de l'\'Ecole Normale Sup\'erieure de Lyon, Universit\'e de Lyon, 46, all\'ee d'Italie, 69007 Lyon, France}
\author{Denis Bartolo}
\affiliation{Laboratoire de Physique de l'\'Ecole Normale Sup\'erieure de Lyon, Universit\'e de Lyon, 46, all\'ee d'Italie, 69007 Lyon, France}

\begin{abstract}
When animal groups move coherently in the form of a flock, their trajectories are not all parallel, the individuals exchange their position in the group. In this Letter we introduce a measure of this mixing dynamics, which we quantify as the winding of the braid formed from the particle trajectories. Building on a paradigmatic flocking model we numerically and theoretically explain the winding statistics, and show that it is predominantly set by the global twist of the trajectories as a consequence of a spontaneous symmetry breaking. 
\end{abstract}

\maketitle
The collective behaviors observed in animal groups have attracted much attention  in the biology,  the mathematics and the physics communities over the last 20 years. Quantitative data analysis have established that the salient traits of collective motion are very well captured by the dynamics of flying spins: persistent random walkers endowed with  interactions akin to ferromagnetic couplings between their velocities~\cite{Vicsek_review,Marchetti_review,Giardina_review,Couzin_review,Theraulaz,Turner}. This framework has been extensively exploited to rationalize structural and dynamical {properties} starting from the  emergence of directed motion, to  rapid (orientational) information transfer, see~\cite{Vicsek1995,ChatePRL,Couzin_review,Cavagna2010,Bialek2012,GiardinaNaturePhys} and references therein. 
However, beyond these spectacular results, the internal dynamics of a flock, the relative motion of the individuals, remains scarcely investigated both experimentally and theoretically yet it is known to display non-trivial anomalous behaviors~\cite{Toner,Cavagna2013}. 

In this Letter, we theoretically describe the mixing statistics of an archetypal flying-spin model.  We first stress the intrinsic geometrical nature of the dynamics of particles in a flock and map this problem to the braiding statistics of their trajectories. We evidence the nontrivial statistics of the winding between pairs of motile-particle trajectories, which is a robust measure of their entanglement. This quantity displays spatial correlations at the population scale. We single out the reason for the nontrivial statistics and show that the spontaneous breaking of a rotational symmetry causes the global twist of the flock to chiefly rule  the long-time winding fluctuations.

{\em Numerical flocking model.}
{ We build on a standard flocking model used to model compact groups akin to those observed in the wild~\cite{Vicsek_review,Couzin_review,ChatePRL,BialekPNAS}. $N$ persistent random walkers, ${\bf r}_i(t)$, $i=1\ldots N$, propel at a constant speed $v_0 = 1$. The dynamics of their orientation $\hat{\vec p}_i(t)$ generically takes the form:
\begin{equation}
\label{EM}
	\frac{\d \hat{\vec p}_i}{\d t} = (\mathbb I - \hat{\vec p}_i \hat{\vec p}_i) \cdot \vec F_i(\{ \vec r_j, \hat{\vec p}_j \}_j) + \boldsymbol \xi_i(t) .
\end{equation}
Rotational diffusion is accounted for by the uncorrelated Gaussian white noises $\boldsymbol \xi_i$ of variance $2D$. The interactions between self-propelled particles amounts to an effective torque which aligns the orientation $\hat{\vec p}_i$ in the direction of $\vec F_i$. The projection operator $(\mathbb I - \hat{\vec p}_i \hat{\vec p}_i)$ ensures that $\hat{\vec p}_i$ lives on the unit sphere. We use a standard (metric) form of ${\bf F}_i$  to account for the flocking dynamics. Noting $\vec r_{ij} \equiv \vec r_i - \vec r_j$ it reads:}
\begin{equation}
\label{interactions}
	\vec F_i = \frac{1}{\tau N_i^A} \sum_{j \in \mathcal A_i} \hat{\vec p}_j + \frac{1}{\tau N_i^B} \sum_{j \in \mathcal B_i} f(r_{ij}) \, \hat{\vec r}_{ij}.
\end{equation}
$\tau$ is a relaxation time, which we henceforth set to $\tau = 1$. The first term in Eq.~\eqref{interactions} is a ferromagnetic term which promotes alignment with the mean direction of the $N_i^A$ neighbors lying in the sphere $\mathcal A_i$ of radius $R_A = 1$. The second term corresponds to attractive and repulsive interactions within the sphere $\mathcal B_i$ of radius $R_B = 5$, it is introduced to yield compact flocks~\cite{Couzin2002}. Following~\cite{ChatePRL,ChateEPJB}, we assume that these interactions are attractive above a distance $2 r_c$, and repulsive when $r_c \leq r_{ij} < 2 r_c$: $f(r_{ij}) = 1 - [r_c/(r_{ij} - r_c)]^5$ with $r_c = 0.4$. Eqs.~\eqref{EM} and~\eqref{interactions} are solved numerically using an explicit Euler scheme.
{ In all that follows, our only control parameter is the noise amplitude, $D$ (see~\cite{Couzin2002,ChatePRL} for a comprehensive investigation of the phase behavior of this model). Below a critical amplitude $D_c \sim 0.3$, the rotational symmetry of the particle orientation is spontaneously broken and collective motion emerges in the form of a flocking transition (see the Supplementary Document~\cite{SI_doc}). A compact polar flock forms and moves along ${\boldsymbol \Pi}(t)=\langle \hat {\bf p}_i(t)\rangle_i$ as exemplified in Fig.~\ref{fig1}(a) and in the supplementary video S1~\cite{SI_movies}. We restrict ourselves to this symmetry-broken phase and investigate the relative motion of the individuals within the flock.}

\begin{figure*}
\begin{center}
\includegraphics[width=\textwidth]{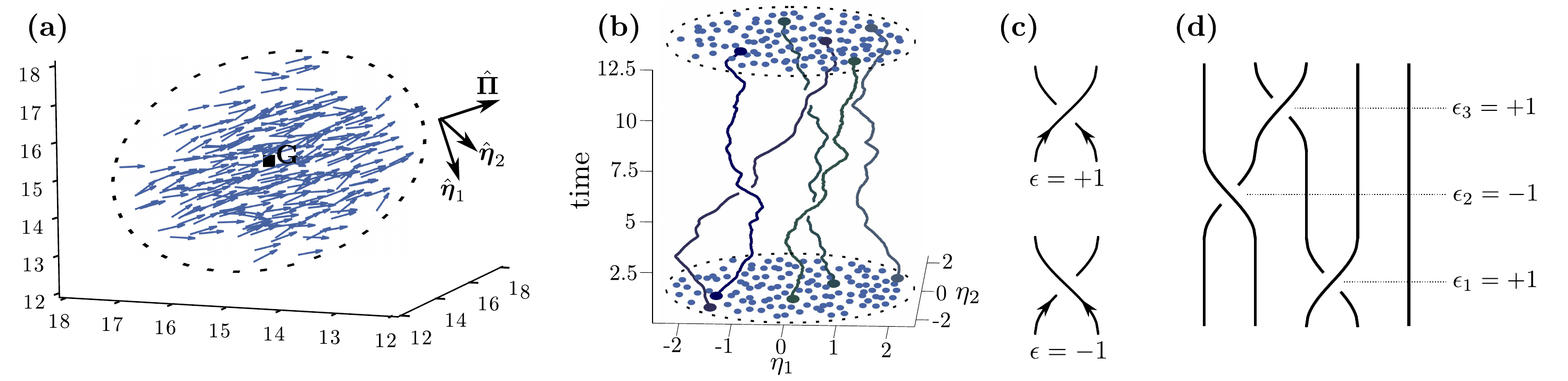}
\caption{(a)~Instantaneous positions and orientations of the particles in a compact polar  flock (250 particles), and definition of the parallel-transported frame $(\hat{\boldsymbol \eta}_1,\hat{\boldsymbol \eta}_2,\hat{\boldsymbol \Pi})$. (b)~World lines of  5 particles in the same polar flock. $D=2.6\times10^{-2}$. (c)~Definition of the crossing index. (d)~Braid diagram associated to the world lines drawn in (b).}
\label{fig1}
\end{center}
\end{figure*}

\paragraph{Quantifying the entanglement of the trajectories.}
{ We quantify the mixing dynamics of the flock by exploiting a powerful toolbox that has been introduced in the context of Lagrangian mixing in fluids. The idea is to relate the mixing  of an ensemble of moving particle  to the entanglement of their wordlines~\cite{Pieranski1996,Boyland2003,Thiffeault2010,Puckett2012}. A convenient measure of the entanglement of a bundle of trajectories is provided by a braid representation that we introduce below.}
{Let us first  define a convenient representation which disentangles the internal dynamics of the flock from the global turns of its mean direction of motion.
We consider an orthogonal frame $(G(t), \hat{\boldsymbol \eta}_1(t),\hat{\boldsymbol \eta}_2(t),\hat{\boldsymbol \Pi}(t))$ shown in Fig.~\ref{fig1}(a). The origin is the center of mass $G(t)$ of the flock. $\hat{\boldsymbol \eta}_1(t=0)$ is chosen arbitrarily;  the 
$\hat{\boldsymbol \eta}_i$s
 are then parallel-transported along the trajectory of $G$ in the course of the dynamics. If the flock were undergoing a rigid-body motion, the particle positions in this frame would be stationary. Conversely, mixing in the flock translates into the winding of the particle  positions in $(\hat{\boldsymbol \eta}_1,\hat{\boldsymbol \eta}_2)$ plane, see Fig.~\ref{fig1}(b) and supplementary movie S2~\cite{SI_movies}.  In order to quantify this winding dynamics, we observe the relative positions of the particles along a reference axis, say $\hat{\boldsymbol \eta}_1$. Particle winding  is measured by observing the particle exchanges along  $\hat{\boldsymbol \eta}_1$, and by assigning an index $\epsilon = \pm 1$ to each crossing, as depicted in Figs.~\ref{fig1}(b),~\ref{fig1}(c) and supplementary movie S3~\cite{SI_movies}. The sign of this index reflects the relative positions of the particle in the orthogonal direction $\hat{\boldsymbol \eta}_2$ as they cross along $\hat{\boldsymbol \eta}_1$. More quantitatively, considering two particles $i$ and $j$, we  introduce their {\it pair winding number}, $w_{ij}(T)$, as the linking number between their world lines: $w_{ij}(T) = \frac{1}{2} \sum_{a} \epsilon_a^{ij}$, where $\epsilon_a^{ij}$ is the index of the $a^{\rm th}$ crossing.  The sum is performed over  the crossings involving the particles $i$ and $j$ only, over a time interval $T$.
This quantity has a clear meaning: it counts the number of turns of particle $i$ around $j$ (or, equivalently, of $j$ around $i$) over a time $T$. The {\it total winding} in the flock between $t_0$ and $t_0+T$ is a measure of the entanglement of the world lines, and hence of the mixing:}
\begin{equation}
\label{defW}
	W(T) =  \frac{1}{\mathcal N_p} \sum_{(i,j)} w_{ij}(T) ,
\end{equation}
where $\mathcal N_p = N(N-1)/2$ is the number of pairs. 
Importantly, $W(T)$ does not depend on the distance between the particles, only the index of the crossings and the times at which they occur matter. Therefore the world lines define a {\it braid} that can be drawn in the form of a normalized braid diagram, Fig.~\ref{fig1}(d)~\cite{Thiffeault2010}. { The total winding corresponds to a topological invariant of the braid: { $W(T) = \sum_a \epsilon_a/(2\mathcal N_p)$}, where we now sum over all the crossings o this simplified representation.}  Practically we choose the Artin representation of the braid word~\cite{Boyland2003,Thiffeault2010,Finn2011}. The {\it braidlab} library~\cite{braidlab} is used to compute both the pair and the total winding numbers. The topological nature of $W(T)$ makes it a very robust measure of the flock mixing. We now carefully investigate its statistics.


\paragraph{Winding statistics.} 
\begin{figure}
\begin{center}
\includegraphics[width=\columnwidth]{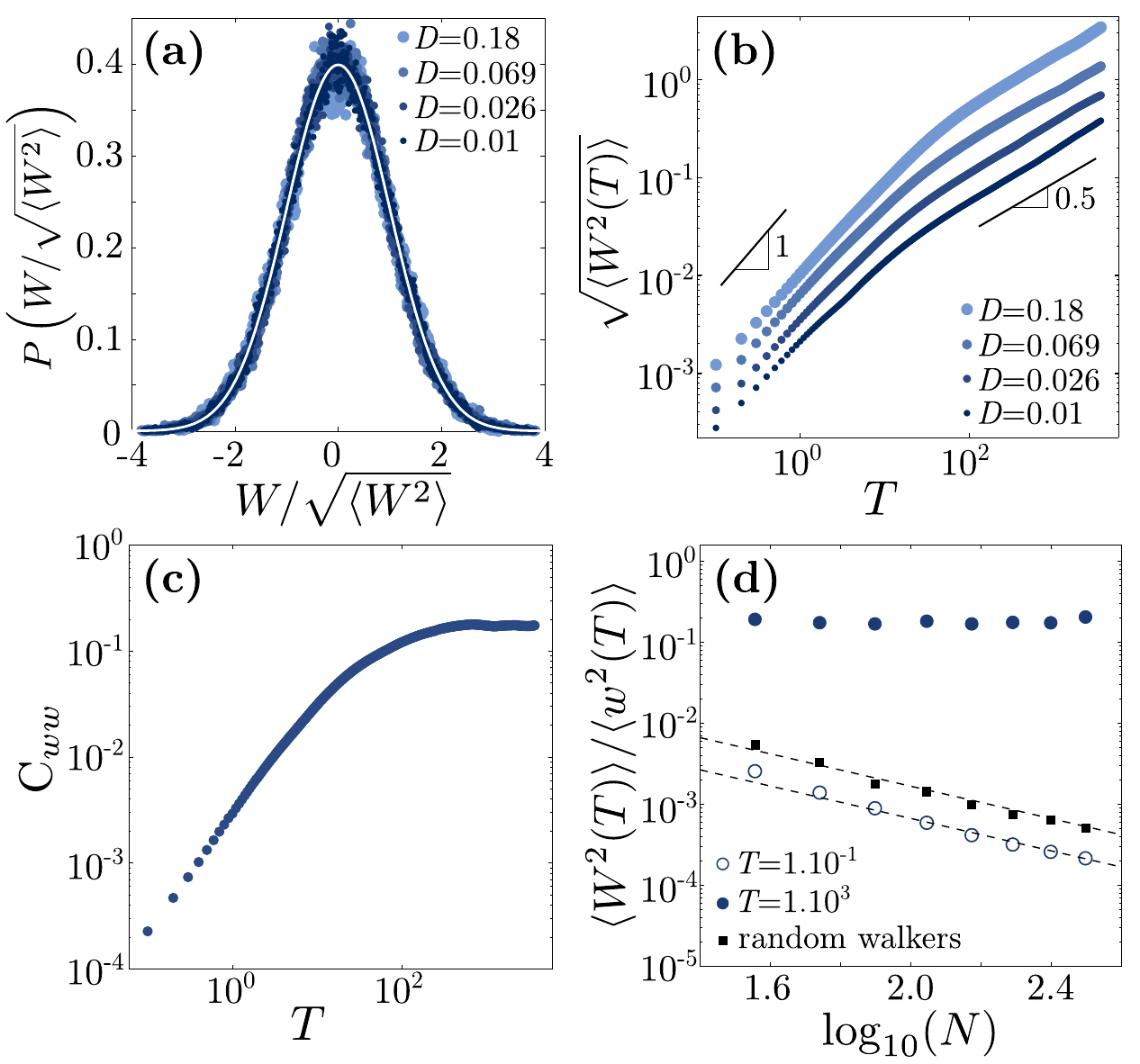}
\caption{(a)~Probability distribution of the total winding $W(T)$, normalized by its standard deviation ($T = 10^3$). The different noise amplitudes correspond to different polarizations of the flock~\cite{SI_doc}. Solid  line: Gaussian distribution. (b)~Standard deviation, ${\langle W^2(T) \rangle}^{1/2}$, as a function of  time $T$, for different noise amplitudes. (c)~Correlation function, defined by Eq.~\eqref{corr}, as a function of $T$. (d)~Variance of the total winding normalized by the variance of the pair winding, plotted versus the particle number. Open circles: $T = 10^{-1}$. Filled circles: $T = 10^3$. Black squares: random walkers confined in a circular box of radius 20, diffusivity: 10. Time step: 0.1. Dashed lines: slope~$-1$.}
\label{fig2}
\end{center}
\end{figure}
The normalized distributions of $W(T)$ are plotted in Fig.~\ref{fig2}(a) for different values of the noise amplitude.  For all the trajectory lengths, the total winding follows a Gaussian statistics with zero mean since the flock has no intrinsic chirality: clockwise and counter-clockwise windings are equally probable. The winding distribution is  fully characterized by its standard deviation ${\langle W^2(T) \rangle}^{\frac{1}{2}}$, where the brackets denote the average over different initial conditions. The winding fluctuations increase linearly with the curvilinear length of the trajectories at short times: ${\langle W^2(T) \rangle}^{\frac{1}{2}} \propto T$, and crosses over to a diffusive regime  where ${\langle W^2(T) \rangle}^{\frac{1}{2}} \propto {T}^{\frac{1}{2}}$ at long times, Fig.~\ref{fig2}(b).
These first results would naively suggest a simple scenario.  If the crossing events were uncorrelated the Gaussian statistics would readily stem from the central limit theorem as $W(T)$ is the average of the crossing signs. In addition, the variance of $W(T)$ would obviously grow in a diffusive manner.
However, this appealing explanation is inconsistent with a deeper analysis of the data. Let us carefully study the correlations between the pair windings, which are quantified by:
\begin{equation}
\label{corr}
	\mathrm{C}_{ww}(T) = \frac{1}{\mathcal N_p (\mathcal N_p-1)} \sum_{(i,j)} \sum_{(k,l) \neq (i,j)} \frac{\langle w_{ij}(t) w_{kl}(T) \rangle}{\langle w^2(T) \rangle} ,
\end{equation}
where $\langle w^2(t) \rangle = \mathcal N_p^{-1} \sum_{(i,j)} \langle w_{ij}^2(T) \rangle$ is the variance of the pair windings. Given this definition,  $\mathrm{C}_{ww}(T) = 0$ for uncorrelated $w_{ij}$s, and $\mathrm{C}_{ww}(T) = 1$ when the pair windings are fully correlated.  Unexpectedly, we find that the correlation between the pair windings does not vanish at long times. Conversely, it increases and  plateaus at a finite value,  Fig.~\ref{fig2}(c), thereby ruling out the simple scenario sketched above. In order to check that this unexpected behavior is not a finite-size artifact, we first note that  $\mathrm{C}_{ww}(T) \sim \langle W^2(T) \rangle / \langle w^2(T) \rangle$ in the large-$\mathcal N_p$ limit and  plot the ratio $\langle W^2(T) \rangle / \langle w^2(T) \rangle$ for flocks of different sizes $N$, in Fig.~\ref{fig2}(d). Whereas short trajectories have indeed winding correlations $\mathrm{C}_{ww}(T)$ that decay with the system size, as $1/N$, the winding correlations of the long trajectories do not display any significant variations with $N$ when increasing the particle number by a factor of $\sim 8$. 
In order to gain more insight into these tho opposite behaviors, we computed the same quantity for the wordlines of  independent 2D  random walkers confined in a circular box. $\mathrm C_{ww}$ follows the same $1/N$ nontrivial scaling observed for short flocking trajectories. We shall note that the finite time step of our numerical scheme regularizes the winding statistics of the random walkers and makes it possible to define its variance~\cite{Belisle1991}. This second set of observations confirms that the saturation of $\mathrm C_{ww}(T)$ at long time originates from extended correlations of the crossing events.

\paragraph{A twist in the statistics.}
We now elucidate the physical mechanism responsible for the winding correlations. We first note that a global instantaneous rotation of the flock around $\hat{\vec \Pi}$ would result in a fully correlated pair winding. We therefore separate the associated global trajectory twist from the total winding. Denoting by $\boldsymbol \eta_i$ the position of particle $i$ projected in the observation plane $(G, \hat{\boldsymbol \eta}_1,\hat{\boldsymbol \eta}_2)$, the instantaneous rotation rate of the flock is:
\begin{equation}
\label{Omega}
	\Omega = \frac{1}{N} \sum_i \frac{1}{\eta_i^2} (\boldsymbol \eta_i \times \dot{\boldsymbol \eta}_i) \cdot \hat{\boldsymbol \Pi} .
\end{equation}
Integrating over time, we define the global twist that is the number of turns of the flock around $\hat{\vec \Pi}$:
\begin{equation}
\label{twist}
	\mathrm{Tw}(T) = \frac{1}{2 \pi} \int_{t_0}^{t_0 + T} \d t \; \Omega(t) .
\end{equation}
We finally define the winding of the untwisted trajectories  in the frame $(G, \hat{\boldsymbol \eta}_1^\star,\hat{\boldsymbol \eta}_2^\star)$, obtained by rotating the parallel-transported frame $(G, \hat{\boldsymbol \eta}_1,\hat{\boldsymbol \eta}_2)$ by an angle $2\pi\times\mathrm{Tw}$:
\begin{equation}
\label{defWstar}
	W^\star = W - \mathrm{Tw} .
\end{equation}
Hence $W^\star =  \mathcal N_p^{-1} \sum_{(i,j)} w_{ij}^\star$, where $w_{ij}^\star = w_{ij} - \mathrm{Tw}$ is the winding between particles $i$ and $j$ in this rotating frame. Within the braid picture, $W^\star$ is found by factorizing out the global twist of the braid word~\cite{Skjeltorp1999,Dehornoy}.

\begin{figure}
\begin{center}
\includegraphics[width=\columnwidth]{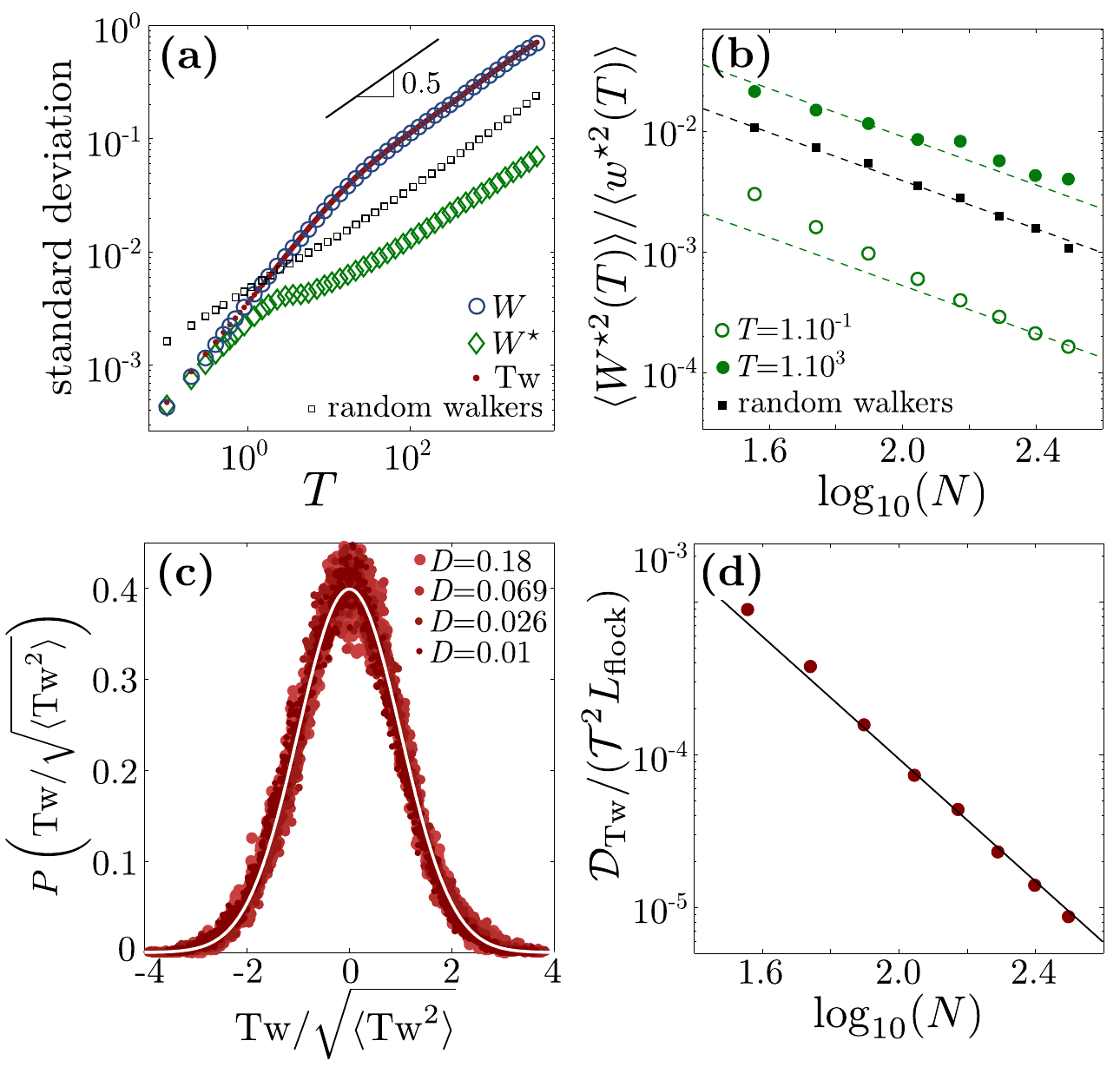}
\caption{(a)~Standard deviations as a function of $T$: ${\langle W^2(T) \rangle}^{1/2}$ (open circles), ${\langle W^{\star 2}(T) \rangle}^{1/2}$ (diamonds) and ${\langle \mathrm{Tw}^2(T) \rangle}^{1/2}$ (filled circles), ${\langle W^2(T) \rangle}^{1/2}$  for confined random walkers (squares), same parameters as in Fig.~2. (b)~Variance of the total winding normalized by the variance of the pair winding computed in the twisting frame for different particle numbers. Open circles: $T = 10^{-1}$. Filled circles: $T = 10^3$. Black squares: confined random walkers. Dashed lines: slope~$-1$. (c)~Probability distribution of the twist normalized by its standard deviation ($T=10^3$). Solid  line: Gaussian distribution. (d)~The normalized twist diffusivity: $\mathcal D_{\rm Tw} / (\mathcal T^2 L_{\rm flock})$ decays quadratically with $N$. Solid line: slope~$-2$.}
\label{fig3}
\end{center}
\end{figure}

In Fig.~\ref{fig3}(a), we plot  the standard deviation of $W$, $\mathrm{Tw}$ and $W^\star$ versus $T$.The twist contribution dominates the total winding at long times and $\langle W^2 \rangle \sim\langle \mathrm{Tw}^2 \rangle$. This numerical fact is better understood by noting that the winding in the rotating frame qualitatively follows the behavior displayed by confined random walkers. Above the relaxation time $\tau = 1$ introduced in Eq.~\eqref{interactions}, $W^\star(T)$ follows the same diffusive evolution: $\langle W^{\star 2}(T) \rangle \propto T$~\cite{Grosberg2003}. More importantly, we also find the same $1/N$ asymptotic scaling for  $\mathrm C_{w^\star w^\star}\sim\langle W^{\star 2}(T) \rangle / \langle w^{\star 2}(T) \rangle$ showing that the spatial correlations of the displacements are short-ranged {\em at all times},  Fig.~\ref{fig3}(b). This result contrasts with the behavior of the total winding in the parallel-transported frame where  $\mathrm C_{w w}\sim\langle W^2(T) \rangle / \langle w^2(T) \rangle$ hardly depends on $N$ at long times, Fig.~\ref{fig2}(d). We can therefore propose the following scenario: the long-range spatial correlations in the flock arise from the global rotation of the flock, hence from the global twist of the trajectories as clearly exemplified in the supplementary video S4~\cite{SI_movies}. 

\paragraph{Hydrodynamic description of the  trajectories' twist.}
The very origin of the global rotation of the flock roots from the spontaneous breaking of the rotational symmetry of particle velocities. This symmetry breaking gives rise to a soft orientational mode~\cite{Bialek2014} which twists  the trajectories in the parallel-transported frame at the entire-flock scale. We now lay out a more quantitative explanation by switching to an hydrodynamic description of the flock viewed as a  active-fluid drop. We use the conventional hydrodynamic framework first introduced phenomenologically by Toner and Tu~\cite{Toner} and later derived from microscopic theories~\cite{Bertin,Farrell2012,Marchetti_review}. The fluid density and velocity fields are $\rho(\vec r,t)$ and $\vec v(\vec r,t)$.
We focus on strongly polarized flocks in which all particles follow the same average direction.
The momentum equation linearized around the homogeneously polarized state takes the simple form~\cite{Marchetti_review, Ramaswamy_review}:
\begin{equation}
\label{hydro}
	\partial_t (\rho \vec v) + \lambda (\hat{\boldsymbol \Pi} \cdot \nabla)\rho \vec v = -\nabla P(\vec \rho) + \Gamma \nabla^2 (\rho \vec v) + \vec f ,
\end{equation}
where $P$ is the local pressure, and $\vec f(\vec r,t)$ is a Gaussian white noise with correlation $\langle f_\alpha(\vec r,t) f_\beta(\vec r',t') \rangle = 2 \tilde D \, \delta (\vec r - \vec r') \delta (t-t') \delta_{\alpha \beta}$.
In this continuous limit, the global rotation rate, Eq.~\eqref{Omega}, is given by:
\begin{equation}
\label{Omega_hydro}
	\Omega = \frac{1}{N} \int \d^3 \vec r \; \frac{1}{\eta^2} (\boldsymbol \eta \times \rho \vec v) \cdot \hat{\boldsymbol \Pi}.
\end{equation}
Two comments are in order.
Firstly, deep in the polarized phase, the linearity of Eq.~\eqref{hydro} implies that the momentum fluctuations are Gaussian. After space and time integration, Eqs.~\eqref{Omega_hydro} and~\eqref{twist} imply that the twist also follows a normal distribution, in agreement with our numerical findings reported in Fig.~\ref{fig3}(c).
Secondly, the damping of velocity fluctuations is set by the diffusive term $\Gamma \nabla^2(\rho \vec v)$, in Eq.~\eqref{hydro} (see~\cite{Marchetti_review, Ramaswamy_review} for more details). Hence the Fourier mode with wave-vector $\vec q$ decays in a time $\sim (\Gamma q^2)^{-1}$. While small-wavelength perturbations are quickly damped, the large-scale fluctuations that occur at the size of the flock, $q \sim 1/L_{\rm flock}$, remain correlated over a time $\mathcal T \sim L_{\rm flock}^2/\Gamma$. This observation  explains the time behavior of the trajectories' twist fluctuations. At short times, $T < \mathcal T$, the small-$q$ fluctuations result in finite-time correlations in the rotation rate $\Omega$. Consequently,  the twist fluctuations persist and undergo a ``balistic'' growth: $\langle \mathrm{Tw}^2(T) \rangle \propto T^2$. At long times, $T > \mathcal T$, all the  Fourier modes  have been relaxed, the correlations vanish, and one recovers the observed diffusive behavior for the trajectory twist: $\langle \mathrm{Tw}^2(T) \rangle \sim \mathcal D_{\rm Tw} T$, Fig.~(3)a.
{  The scaling of the effective diffusivity with the size of the flock is computed in~\cite{SI_doc}: $\mathcal D_{\rm Tw} \propto \tilde D \mathcal T^2 L_{\rm flock} N^{-2}$. This prediction  agrees again with our numerical observations, Fig.~\ref{fig3}(d). Now that we have elucidated the twist statistics, we finally explain why it dominates the winding  in large flocks. Assuming that the flock density weakly depends on $N$, we find that $\langle \mathrm{Tw}^2(T) \rangle\sim N^{-1/3}$. As  $\langle W^{\star 2}\rangle/\langle w^{\star 2}\rangle\sim N^{-1}$, Fig.~\ref{fig3}(b), we conclude that the ratio between  the local winding fluctuations  and the twist vanishes as $N\to\infty$.

In order to unambiguously prove that the soft rotational mode chiefly rules the winding statistics, 
we stress that all this phenomenology is lost in isotropic swarms. We show in the Supplementary Document~\cite{SI_doc} that when the noise amplitude is too strong to observe directed motion,  $\langle W^2 \rangle$ deviates from $\langle {\rm Tw}^2 \rangle$: increasing the noise results in the decorrelation  of the winding and the twist fluctuations~\cite{SI_doc}.
Altogether, our numerical and analytical results  suggest a strong robustness of our main findings. The prominence of the twist fluctuations, leading to a coherent braiding of the flock trajectories, is expected to be qualitatively robust to the very details of the  interactions. It solely relies on the spontaneous breaking of the rotational symmetry, and on its associated soft mode. It is therefore a direct consequence of the flocking transition and should be observed in all models yielding polarized flocks. }
 
 { We shall close with Letter from an experimental perspective. In most of the situations, in the wild, external perturbations and fields  explicitly break the rotational symmetry (e.g. gravity, predators, obstacles), or cause sudden collective turns as analyzed in~\cite{Cavagna2013,Attanasi2014a,Attanasi2014b,Cavagna2014} for starling flocks. The winding statistics should be an effective probe of the flock response to external bias without any a priori knowledge about the individual propulsion and interaction mechanisms.}
 
%
%

\begin{acknowledgments}
We acknowledge support from Institut Universitaire de France and ANR project MiTra.
\end{acknowledgments}


\balancecolsandclearpage

\onecolumngrid
\appendix

\titleformat{\section}{\center \small \bfseries}{\thesection. }{0pt}{\MakeTextUppercase}   \titlespacing*{\section}{0pt}{6ex plus 1ex minus .2ex}{4ex plus .2ex}
\titleformat{\subsection}{\center \small \bfseries}{\thesubsection. }{0pt}{}    \titlespacing*{\subsection} {0pt}{3.25ex plus 1ex minus .2ex}{3ex plus .2ex}

\begin{center}
{\bf \Large Supplementary Information}
\end{center}
\vspace{0.1cm}

\section{Transition to collective motion}

\begin{figure}[h!]
\begin{center}
\includegraphics[width=0.32\columnwidth]{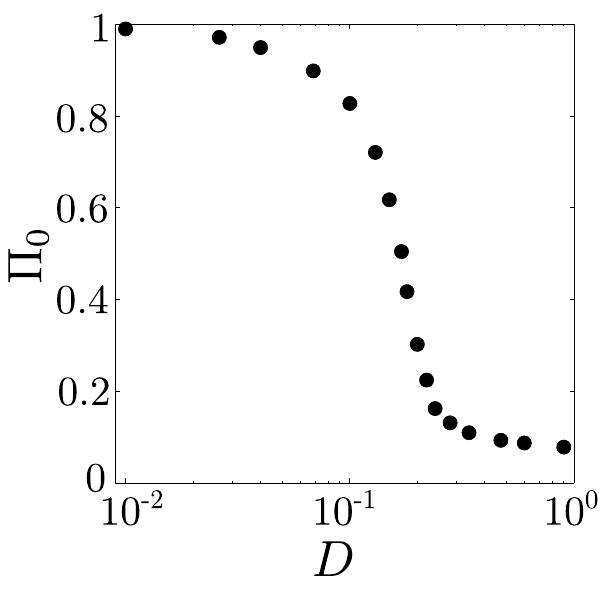}
\caption{Mean polarization of the flock plotted versus the noise amplitude.}
\label{fig_S1}
\end{center}
\end{figure}

Upon decreasing the noise amplitude, the  model defined by Eqs. (1) and (2) in the main text displays a transition to directed motion, upon decreasing the noise amplitude below a critical value $D_c \sim 0.2$. The mean polarization, $\Pi_0 = N^{-1} \left\vert \sum_i \hat{\vec p}_i \right\vert$, increases from 0 in isotropic flocks to 1 in coherenty-moving groups, as shown in Fig.~\ref{fig_S1}. Both polar and isotropic flocks are compact: due to the attractive interactions, they do not span the entire simulation box but keep a finite size. A comprehensive characterization of the nature of the transition is provided in~\cite{ChatePRL}.

\section{Long-time behavior of the twist fluctuations in polar flocks}

We detail the derivation of the scaling law for the twist diffusivity $\mathcal D_{\rm Tw}$ from the hydrodynamic description of the flock. For the sake of clarity, we introduce the momentum field $\tilde{\vec V} = \rho \vec v$. As we noted in the main text, its Fourier component $\tilde{\vec V}_{\vec q}$, with wave-vector $\vec q$, is damped over a typical time $\sim (\Gamma q^2)^{-1}$. From Eq.~(8), the time correlations therefore decay as:
\begin{equation}
	\langle \tilde V_{\vec q, \alpha}(t) \tilde V_{\vec q',\beta}(t') \rangle \propto \frac{\tilde D}{\Gamma q^2} {\rm e}^{-\Gamma q^2 \vert t-t' \vert} \, \delta_{\alpha \beta} \, \delta(\vec q + \vec q') ,
\end{equation}
where the indices $\alpha$, $\beta$ denote the spatial components of $\tilde{\vec V}_{\vec q}$. The smallest wave-vector being $q_{\rm min}\sim 1/L_{\rm flock}$, the time correlations vanish is the limit where $\vert t-t' \vert \gg \mathcal T \sim L_{\rm flock}^2/\Gamma$. In this regime, the momentum field is delta-correlated:
\begin{equation}
\label{correlations_W}
	\langle \tilde V_{\vec q, \alpha}(t) \tilde V_{\vec q',\beta}(t') \rangle \propto \frac{\tilde D}{\Gamma^2 q^4} \delta(t-t') \, \delta_{\alpha \beta} \, \delta(\vec q + \vec q') .
\end{equation}
From this result, we deduce the autocorrelation of the global rotation rate. Eq.~(9) provides the following expression:
\begin{equation}
	\langle \Omega(t) \Omega(t') \rangle = \frac{1}{N^2} \int \d^3 \vec r \int \d^3 \vec r' \frac{1}{\eta^2 \eta'^2} \hat{\Pi}_\alpha \hat{\Pi}_\lambda\epsilon_{\alpha \beta \gamma} \epsilon_{\lambda \mu \nu} \eta_\beta \eta_\mu' \; \langle \tilde V_\gamma(t) \, \tilde V_\nu(t') \rangle ,
\end{equation}
where we recall that $\boldsymbol \eta = (\mathbb I - \hat{\boldsymbol \Pi} \hat{\boldsymbol \Pi})\cdot \vec r$ and $\epsilon_{\alpha \beta \gamma}$ is the fully antisymmetric Levi-Civita symbol. Moving to Fourier space and using Eq.~\eqref{correlations_W}, we have equivalently:
\begin{equation}
	\langle \Omega(t) \Omega(t') \rangle = \frac{\tilde D}{\Gamma^2 N^2} \int \d^3 \vec r \int \d^3 \vec r' \int \d^3 \vec q \, \frac{\boldsymbol \eta \cdot \boldsymbol \eta'}{\eta^2 \eta'^2} \frac{1}{q^4}  {\rm e}^{-i \vec q \cdot(\vec r - \vec r')} \;  \delta(t-t').
\label{dimensionalanalysis}
\end{equation}
We therefore obtain $\langle \Omega(t) \Omega(t') \rangle \sim \mathcal D_{\rm Tw} \, \delta(t-t')$. A simple dimensional estimate of the value of the integral in Eq.~\ref{dimensionalanalysis} readily provides the scaling $\mathcal D_{\rm Tw} \propto \tilde D \Gamma^{-2} N^{-2} L_{\rm flock}^5$ for the effective diffusivity. Recalling that the persistence time of the twist is $\mathcal T \sim L_{\rm flock}^2/\Gamma$, we recover the scaling given in the main text:
\begin{equation}
	\mathcal D_{\rm Tw} \propto \tilde D \mathcal T^2 L_{\rm flock} N^{-2} .
\end{equation}

\section{Winding statistics across the transition to collective motion}

\begin{figure}[h!]
\begin{center}
\includegraphics[width=0.62\columnwidth]{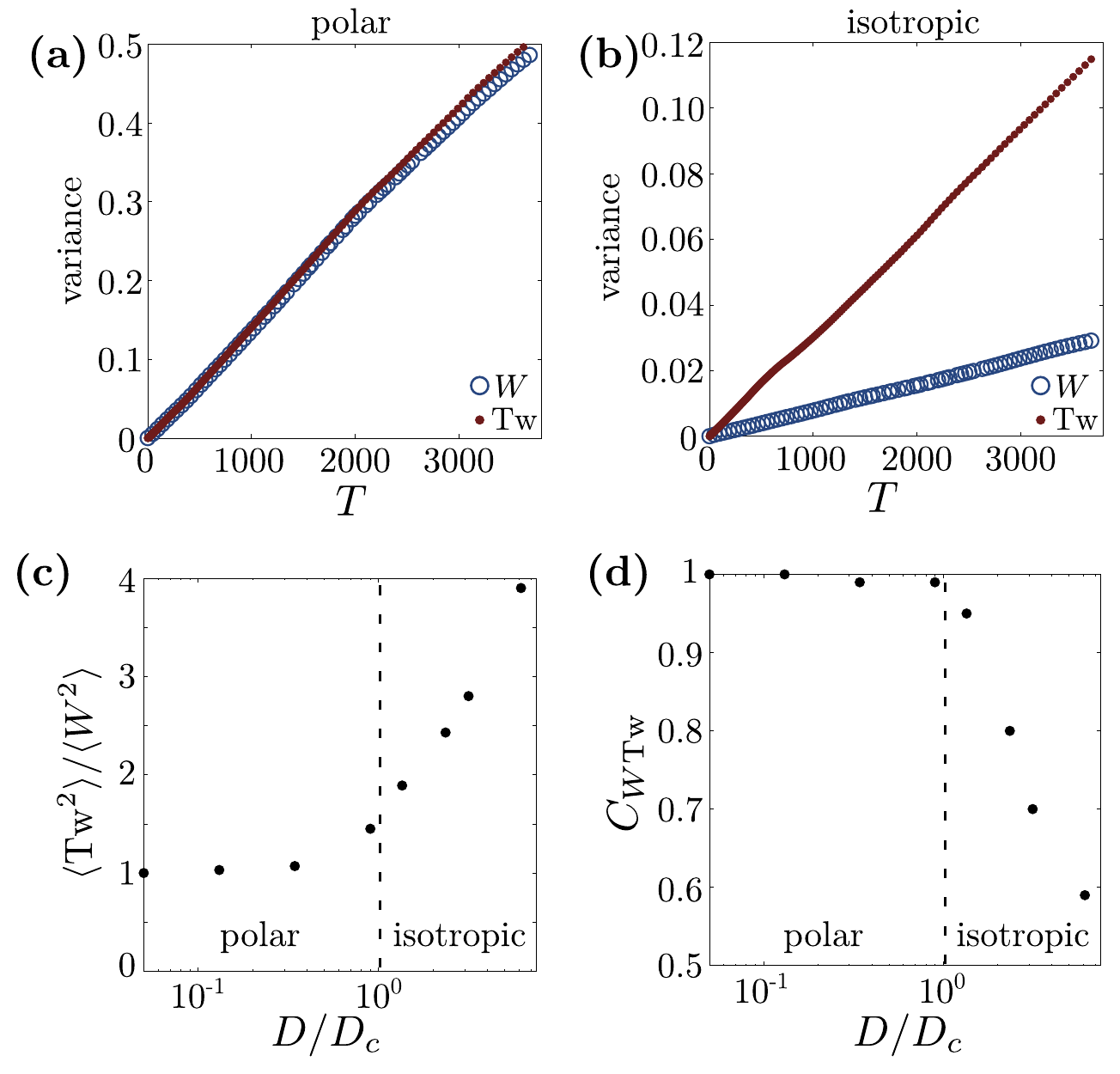}
\caption{(a)~Variance of the total winding $\langle W^2(T)$ (blue open circles), and of the twist $\langle {\rm Tw}^2(T)$ (red filled circles), plotted as a function of time $T$ for a polarized flock ($D = 0.02$). (b)~Same as~(a) for an isotropic flock ($D = 1.23$). (c)~Ratio between the variances of the twist and of the total winding at long times, plotted versus the noise amplitude. (d)~Long-time correlation between the twist and the total winding, $C_{W {\rm Tw}} = \langle W {\rm Tw} \rangle/[\langle W^2 \rangle \langle {\rm Tw}^2 \rangle]^{1/2}$.}
\label{fig_S2}
\end{center}
\end{figure}

In the main text, we restricted ourselves to polarized flocks. Here, we extend  the analysis to higher noise amplitudes, above the transition to collective motion. As the flock has no well-defined direction of motion, we project the particle positions along a fixed direction (rather than in a parallel-transported frame). We compute the winding and the twist statistics as explained in the main text. We then compare, in Fig.~\ref{fig_S2}, the results obtained for polarized and isotropic flocks. We plot the variances $\langle W^2(T) \rangle$ and $\langle {\rm Tw}^2(T) \rangle$ in two extreme cases, see Figs.~\ref{fig_S2}(a) and~(b). We also quantify the ratio $\langle {\rm Tw}^2 \rangle / \langle W^2 \rangle$ at long times, as well as the correlations between the winding and the twist fluctuations, $C_{W {\rm Tw}}$, in Figs.~\ref{fig_S2}(c) and~(d). Upon varying the noise across the transition to collective motion, the emergence of polar order is associated with a clear change in the winding statistics.
\begin{itemize}
\item In polar flocks (at low noise amplitude, $D < D_c \sim 0.2$), we find that $\langle {\rm Tw}^2 \rangle \sim \langle W^2 \rangle$. The total winding and the twist are fully correlated, $C_{W {\rm Tw}} \sim 1$. As they both follow a Gaussian distribution, these two quantities are statistically equivalent.
\item In isotropic flocks ($D > D_c$), the total winding is well distinct from the twist as exemplified in Fig.~\ref{fig_S2}(b). In contrast with polar flocks, the ratio $\langle {\rm Tw}^2 \rangle / \langle W^2 \rangle$ deviates from unity upon increasing the noise amplitude, see Fig.~\ref{fig_S2}(c). This behavior is associated with a clear decay of the correlation between the winding and the twist in an isotropic flock, Fig.~\ref{fig_S2}(d).
\end{itemize}

These results further confirm the scenario proposed in the main text to explain the winding statistics. In polar flocks, the total winding and the twist are both ruled by the same stochastic process: the spontaneous breaking of the rotational symmetry yields a spatially coherent rotation of the flock yet stochastic in time, i.e.\ a global braiding of the trajectories. This mechanism does not rely on the specific form of the interactions leading to polar order, it is therefore generic to all flocking models. By contrast, in isotropic flocks, this correlated rotational mode is suppressed as rotational symmetry is not broken. The residual winding and twist fluctuations solely arise from the weakly-correlated displacements of the individuals: in isotropic flocks, they correspond to different random processes.

\end{document}